\documentclass[%
 aip,
 jmp,%
 amsmath,amssymb,
reprint,%
]{revtex4-1}
\usepackage{cases}
 \usepackage{subeqnarray}
\usepackage{amsfonts}
\usepackage{cases}

\usepackage{cases}
\usepackage{graphicx}
\usepackage{dcolumn}
\usepackage{bm}
\include{biblio}

\begin{document}

\title{Magnetically-tunable cutoff in asymmetric thin metal film plasmonic waveguide}
\author{Song-Jin Im}
\email{ryongnam31@yahoo.com or sj.im@ryongnamsan.edu.kp}
\affiliation{Department of Physics, Kim Il Sung University, Taesong District, 02-381-4410 Pyongyang, Democratic People's Republic of Korea}
\author{Chol-Song Ri}
\affiliation{Department of Physics, Kim Il Sung University, Taesong District, 02-381-4410 Pyongyang, Democratic People's Republic of Korea}
\author{Ji-Song Pae}
\affiliation{Department of Physics, Kim Il Sung University, Taesong District, 02-381-4410 Pyongyang, Democratic People's Republic of Korea}
\author{Yong-Ha Han}
\affiliation{Department of Physics, Kim Il Sung University, Taesong District,  02-381-4410 Pyongyang, Democratic People's Republic of Korea}
\author{Joachim Herrmann}
\email{jherrman@mbi-berlin.de}
\affiliation{Max-Born-Institute for Nonlinear Optics and Short Pulse Spectroscopy, Max-Born-Str. 2a,
D-12489 Berlin, Germany}
\date{\today}
\begin{abstract}
We theoretically investigated the magnetically-tunable cutoff of long-range surface plasmon polariton along thin metal film surrounded by a magneto-optic material on one side and by a nonmagnetic dielectric on the other side. The analytically derived cutoff condition predicts that a magnetic field bias can induce a novel degenerate cutoff-state near which the beyond-cutoff radiation in one side can be switched to that in the other side by a minor variation of the magnetic field from the bias. The magnetization bias needed for the degeneracy is in proportion to the metal film thickness and in inverse proportion to the wavelength.
\end{abstract}
\pacs{}
\keywords{}

\maketitle

Surface plasmon polaritons are sensitive to the optical properties of the environmental dielectric materials, which is promising for application in sensing technology \cite{Anker2008}. In particular, the long-range surface plasmon polariton mode of thin metal film is highly sensitive to the index-asymmetry of the environmental dielectric materials \cite{Magno2013} showing the cutoff behavior \cite{Sarid1981,Yang1991,Pierre2009} where the long-range surface plasmon polariton mode is changed to a radiation mode in one side layer with the higher refractive index. On the other hand, magnetic field has been demonstrated to be an excellent candidate for active plasmonics \cite{Armelles2013,Temnov2010,Belotelov2011,Temnov2012,Belotelov2013,Mankel2013,Janusonis2016}. The external magnetic field influences surface plasmon polaritons in planar plasmonic structures including ferromagnetic metal \cite{Temnov2010,Temnov2012} or dielectric \cite{Belotelov2011,Belotelov2013} layers. Here, the gyration, which is the magnetically-induced off-diagonal permittivity of the ferromagnetic material, leads to a change of wavenumber of the surface plasmon polaritons keeping the transverse magnetic peculiarity \cite{Temnov2010,Belotelov2011}. Moreover, the recently demonstrated ultrafast magnetization dynamics would enable the switching speed to reach to a femtosecond level \cite{Kimel2005,Stanciu2007,Lambert2014,Cornelissen2016}.

In this paper we theoretically investigate the magnetically-tunable cutoff in asymmetric thin metal film plasmonic waveguide. A magnetic field bias can induce a novel degenerate cutoff-state near which the beyond-cutoff radiation in one side can be switched to that in the other side by a minor variation of the magnetic field from the bias. The minimum magnetization bias needed for the degeneracy is in proportion to the metal film thickness and in inverse proportion to the wavelength.

\begin{figure}
\includegraphics[width=0.5\textwidth]{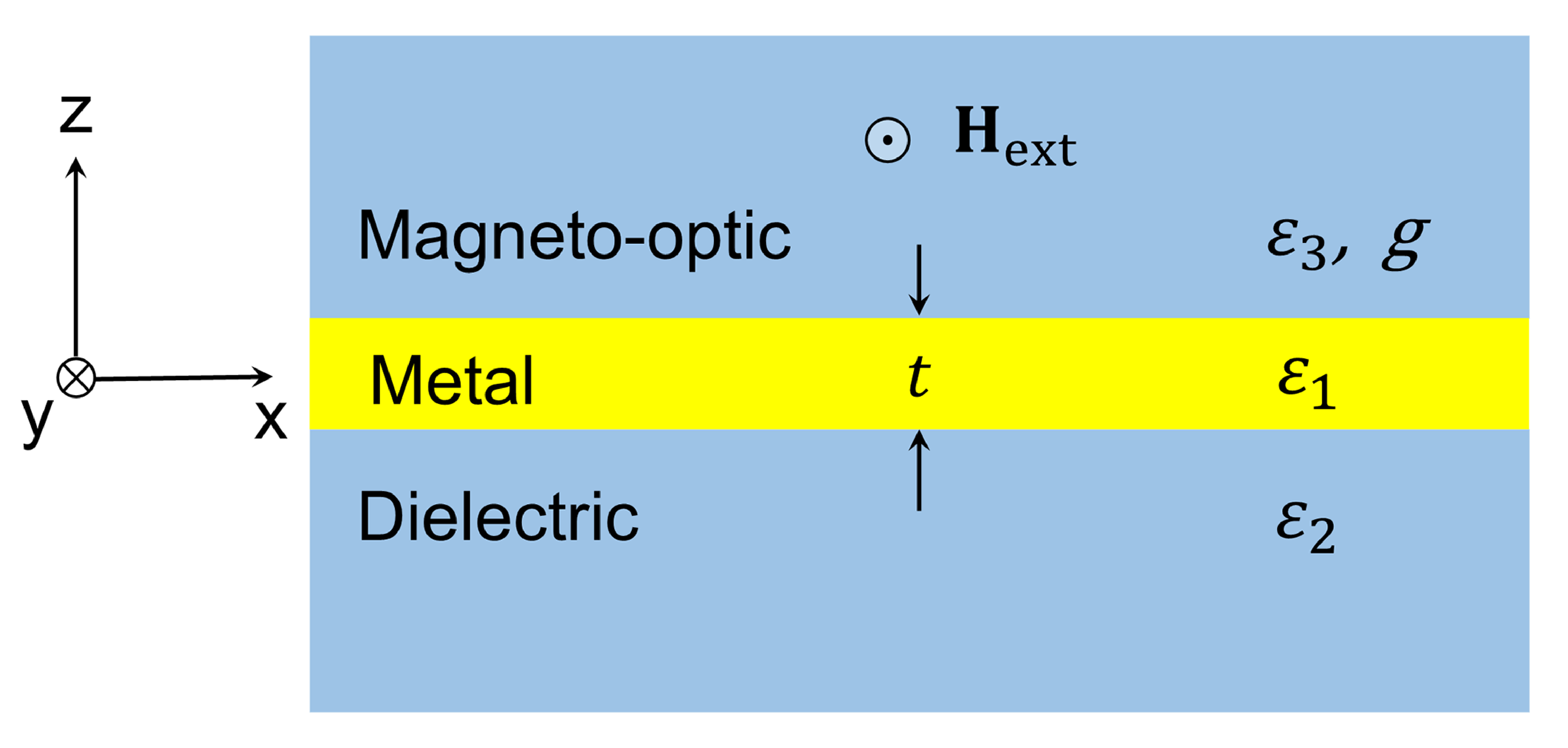}
\caption{Metal film with the thickness $t$ and the permittivity ${\varepsilon _1}$ surrounded on one side by a magneto-optic material with the diagonal permittivity  ${\varepsilon _3}$  and the magnetically-induced gyration $g$  and on the other side by a nonmagnetic dielectric with the permittivity ${\varepsilon _2}$ .}
\label{fig:1}
\end{figure} 

\begin{figure}
\includegraphics[width=0.5\textwidth]{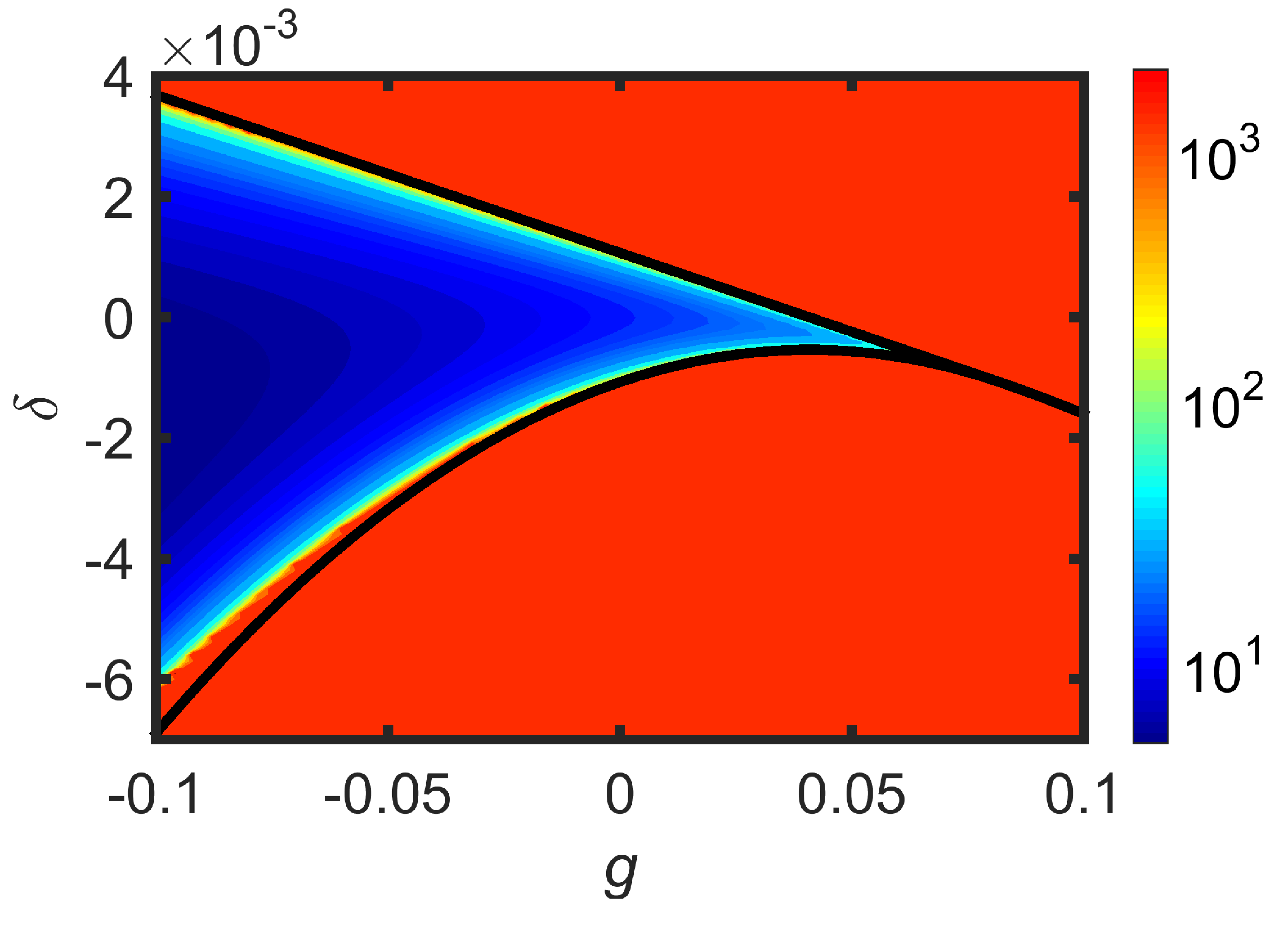}
\caption{Mode extension $L = {L_2} + {L_3}$  in the unit of $\mu$m for the asymmetric thin metal film structure as in Fig.~\ref{fig:1}, where ${L_2} = 1/(2{\mathop{\rm Re}\nolimits} {k_2})$ and  ${L_3} = 1/(2{\mathop{\rm Re}\nolimits} {k_3})$ by numerically mode-solving the Maxwell equation. The black curve is by Eq.~(\ref{eq6}), (\ref{eq8}) and (\ref{eq9}). Here, we assumed   $\lambda$=1550 nm, $t$=5 nm and ${\varepsilon_3}$=2.5.}
\label{fig:2}
\end{figure} 

\begin{figure}
\includegraphics[width=0.5\textwidth]{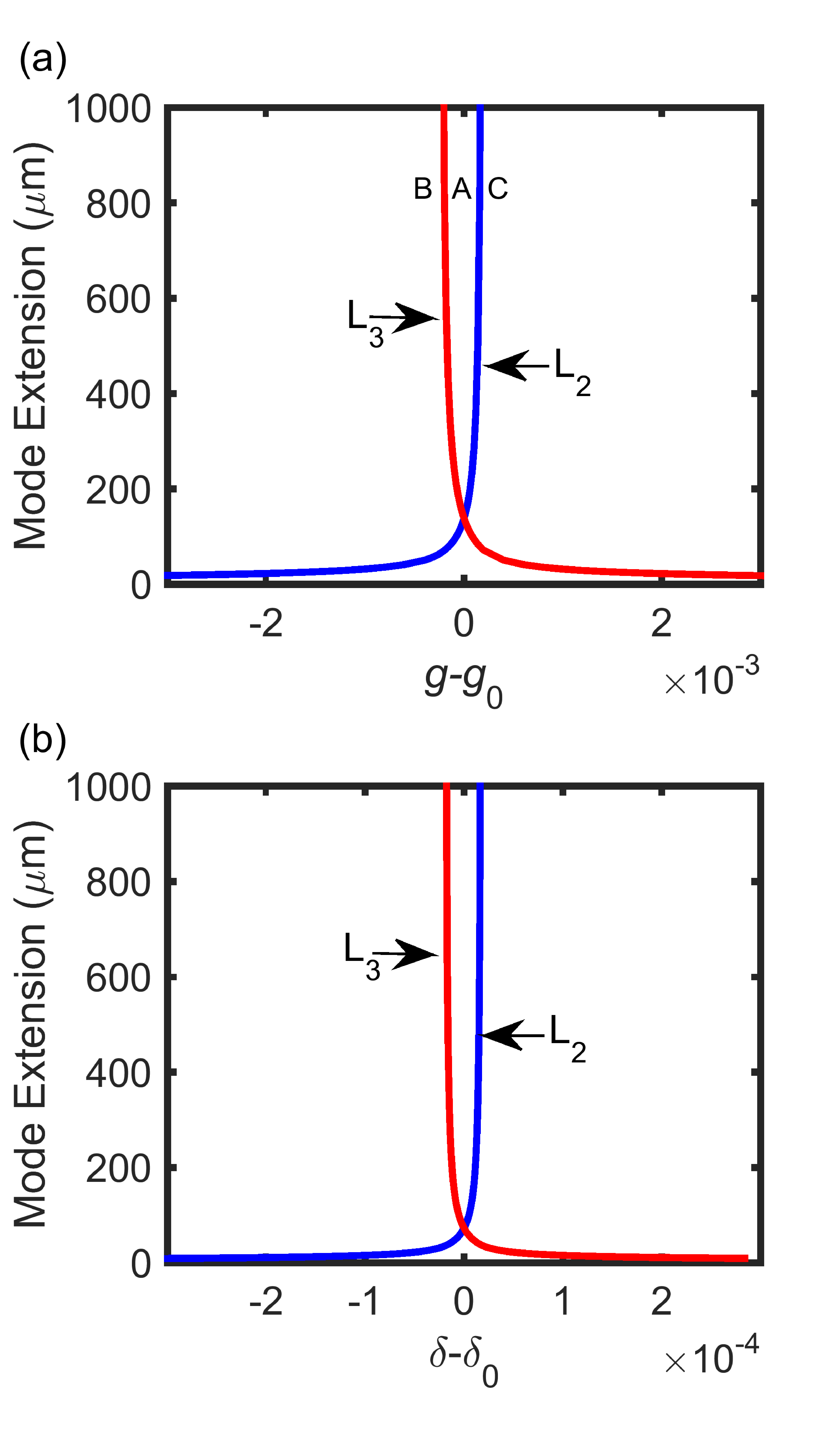}
\caption{Mode extensions to the side of  nonmagnetic dielectric ${L_2}$ (the blue curve) and to the side of  magneto-optic material ${L_3}$ (the red curve) according to the deviation of the gyration $g$ (a) and the index-asymmetry $\delta$ (b) from the degenerate cutoff-state $({g_0},{\delta _0})$ by Eq.~(\ref{eq9}).  The other parameters are same as in Fig.~\ref{fig:2}.}
\label{fig:3}
\end{figure} 

\begin{figure}
\includegraphics[width=0.5\textwidth]{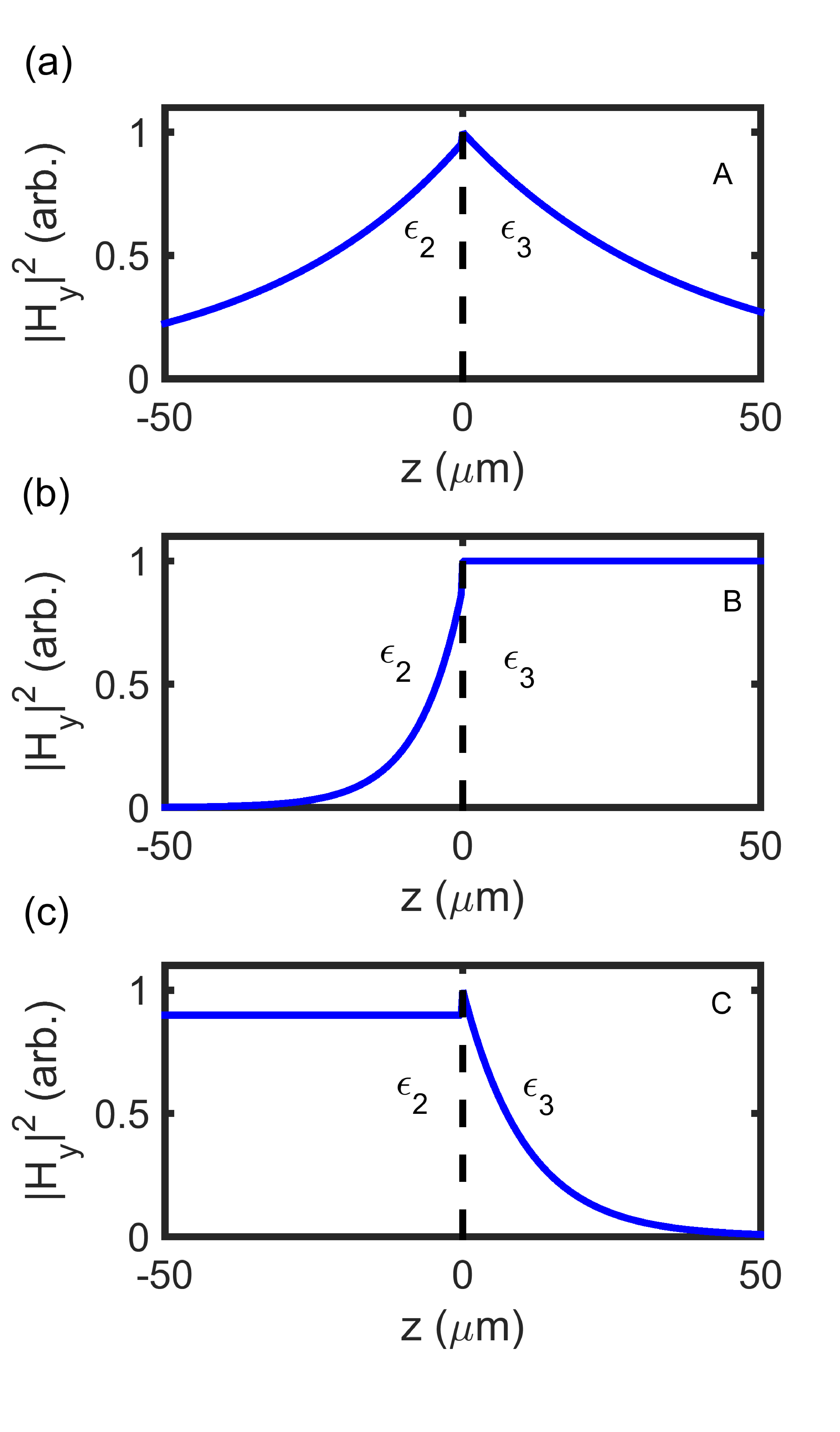}
\caption{Mode distributions for the bound mode (a), the radiation mode in the magneto-optic material layer (b) and the radiation mode in the nonmagnetic dielectric layer (c). The gyration values $g$ of (a), (b) and (c) correspond to the gyration values for A, B and C of Fig.~\ref{fig:3}, respectively. The other parameters are same as in Fig.~\ref{fig:2}.}
\label{fig:4}
\end{figure} 

\begin{figure}
\includegraphics[width=0.5\textwidth]{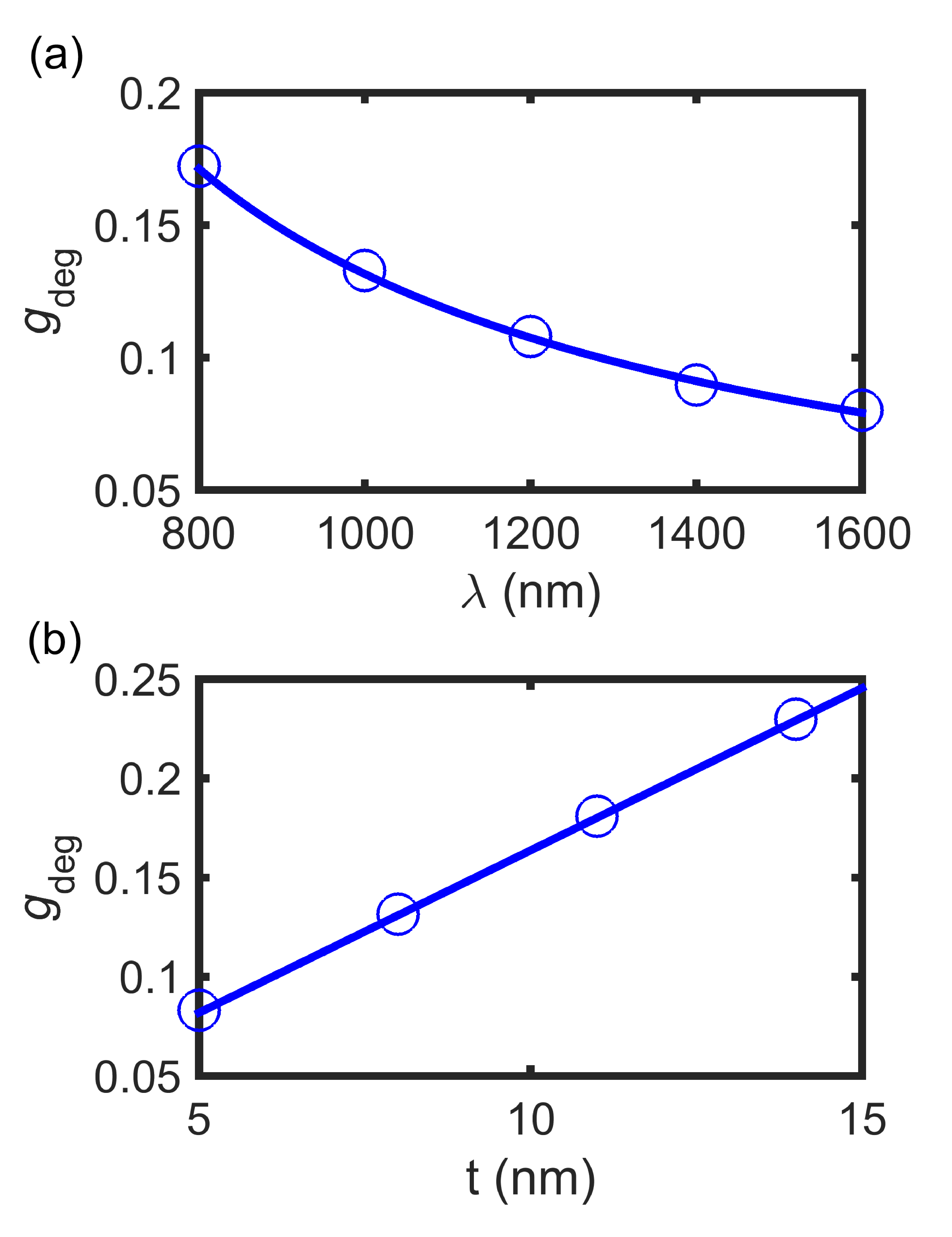}
\caption{Dependence of ${g_{\deg }}$ on the wavelength $\lambda$  and the metal film thickness $t$ . The blue curves are by Eq.~(\ref{eq6b}). The blue circles are by the numerical mode solution results.}
\label{fig:5}
\end{figure} 

The geometry of the thin metal film surrounded on one side by a magneto-optic material and on the other side by a nonmagnetic dielectric is shown in the Fig.~\ref{fig:1}. Surface modes of the thin metal film is described by the Maxwell equation in the case of absence of external charge and current,

\begin{eqnarray}
\nonumber
\nabla  \times {\bf{E}} =  - \frac{{\partial {\bf{{B}}}}}{{\partial t}},\\
\nabla  \times {\bf{H}} = \frac{{\partial {\bf{D}}}}{{\partial t}}.
\label{eq1}
\end{eqnarray}
While an external magnetic field is in the direction of y-axis, the permittivity tensor of the magneto-optic material can be expressed as

\begin{eqnarray}
\hat \varepsilon  = \left( {\begin{array}{*{20}{c}}
{{\varepsilon _3}}&0&{ig}\\
0&{{\varepsilon _3}}&0\\
{ - ig}&0&{{\varepsilon _3}}
\end{array}} \right),
\label{eq2}
\end{eqnarray}
where the gyration $g$ in the off-diagonal components  is proportional to the magnetization. As the off-diagonal permittivity induced by the transverse magnetic field does not change the TM peculiarity of surface plasmon polariton mode\cite{Temnov2010,Belotelov2011}, the TM mode (${E_y} = 0,{H_x} = {H_z} = 0$) is assumed. The magnetic field component can be expressed as following.

\begin{numcases}{{H_y} =}
\nonumber
A{\text{exp}}(i\beta x - {k_3}z), &for $z > t/2$,\\
\nonumber
B{\text{exp}}(i\beta x + {k_2}z), &for $z <  - t/2$,\\
C{\text{exp}}(i\beta x + {k_1}z) + \\
\nonumber
D{\text{exp}}(i\beta x - {k_1}z), &for $- t/2 < z < t/2$.
\label{eq3}
\end{numcases}
From the Maxwell equation and the continuity of ${E_x}$ and ${H_y}$ at the interfaces,
\begin{eqnarray}
{q_1}\left( {{q_2} + {q_3}} \right) + ({q_2}{q_3} + q_1^2)\tanh ({k_1}t) = 0,
\label{eq4}
\end{eqnarray}
where ${q_1} = {k_1}/{\varepsilon _1}$, ${q_2} = {k_2}/{\varepsilon _2}$ and ${q_3} = ({\varepsilon _3}{k_3} + g\beta )/(\varepsilon _3^2 - {g^2})$ and $k_1^2 = {\beta ^2} - k_0^2{\varepsilon _1}$ , $k_2^2 = {\beta ^2} - k_0^2{\varepsilon _2}$  and $k_3^2 = {\beta ^2} - (\varepsilon _3^2 - {g^2})k_0^2/{\varepsilon _3}$. 

In the case of $g=0$ and ${\varepsilon _2} = {\varepsilon _3}$, the symmetric thin metal film structure supports  ${H_y}$-symmetric and -antisymmetric modes. The  ${H_y}$-symmetric mode, which has the smaller attenuation and the weaker confinement, is called the long-range surface plasmon polariton mode \cite{Pierre2009}.  If a small value of the index-asymmetry $\delta  = ({\varepsilon _2} - {\varepsilon _3})/{\varepsilon _3} << 1$  is introduced, the long-range surface plasmon polariton mode is changed to a radiation mode in the side layer with the higher refractive index, which is called the cutoff behavior \cite{Yang1991}. If ${k_1}t <  < 1$   and  $\left| {{\mathop{\rm Im}\nolimits} ({\varepsilon _1})/{\mathop{\rm Re}\nolimits} ({\varepsilon _1})} \right| <  < 1$ are assumed, the cutoff condition is ${\delta _{c2}} =  - {\delta _{c3}} = 4{\pi ^2}{t^2}/{\lambda ^2{(1 - {\varepsilon _3}/{\varepsilon _1})^2}}{\varepsilon _3}$ \cite{Yang1991}. Here,   ${\delta _{c2}}$ and  ${\delta _{c3}}$  are the index-asymmetries corresponding to both cutoff-states for the  ${\varepsilon _2}$-layer radiation and for the  ${\varepsilon _3}$-layer radiation, respectively.

Now we introduce a small gyration $g/{\varepsilon _3} <  < 1$.  At the cutoff-state   $({\delta _{c2}},{g_{c2}})$ for the radiation in the side of the nonmagnetic dielectric, 
\begin{eqnarray}
\nonumber
{q_2} = 0,\\
\nonumber
{q_3} = {k_0}(\sqrt {{\delta _{c2}} + g_{c2}^2/\varepsilon _3^2}  + {g_{c2}}/{\varepsilon _3})/\sqrt {{\varepsilon _3}} ,\\
{q_1} = {k_0}\sqrt {1 - {\varepsilon _3}/{\varepsilon _1}} /\sqrt { - {\varepsilon _1}} .
\label{eq5}
\end{eqnarray}
If we substitute Eq.~(\ref{eq5}) to Eq.~(\ref{eq4}),
\begin{subeqnarray}
\label{eq6}
 \slabel{eq6a}
{\delta _{c2}} = \frac{{{{({g_{\deg }} - {g_{c2}})}^2}}}{{\varepsilon _3^2}} - \frac{{g_{c2}^2}}{{\varepsilon _3^2}}, \,\ \text{for}       \,\,\,{g_{c2}} < {g_{\deg }},\\
 \slabel{eq6b}
{g_{\deg }} = 2\pi \frac{t}{\lambda }{\varepsilon _3}\sqrt {{\varepsilon _3}} (1 - \frac{{{\varepsilon _3}}}{{{\varepsilon _1}}}).
\end{subeqnarray}
At the cutoff-state  $({\delta _{c3}},{g_{c3}})$  for the radiation in the side of the magneto-optic material,
\begin{eqnarray}
\nonumber
{q_2} = {k_0}\sqrt { - ({\delta _{c3}} + g_{c3}^2/\varepsilon _3^2)} /\sqrt {{\varepsilon _3}}, \\
\nonumber
{q_3} = {k_0}({g_{c3}}/{\varepsilon _3})/\sqrt {{\varepsilon _3}}, \\
{q_1} = {k_0}\sqrt {1 - {\varepsilon _3}/{\varepsilon _1}} /\sqrt { - {\varepsilon _1}}.
\label{eq7}
\end{eqnarray}
If we substitute Eq.~(\ref{eq7}) to Eq.~(\ref{eq4}),
\begin{eqnarray}
{\delta _{c3}} =  - \frac{{{{({g_{\deg }} - {g_{c3}})}^2}}}{{\varepsilon _3^2}} - \frac{{g_{c3}^2}}{{\varepsilon _3^2}}, \,\, \text{for} \,\,\,        {g_{c3}} < {g_{\deg }}.
\label{eq8}
\end{eqnarray}
The both cutoff-states $({\delta _{c2}},{g_{c2}})$ and $({\delta _{c3}},{g_{c3}})$ are degenerate with $({\delta _{0}},{g_{0}})$,
\begin{eqnarray}
{\delta _0} =  - \frac{{g_0^2}}{{\varepsilon _3^2}}, \,\, \text{for} \,\,\,   {g_0} \ge {g_{\deg }}.
\label{eq9}
\end{eqnarray}
Fig.~\ref{fig:2} shows the mode extension $L = {L_2} + {L_3}$ in the unit of $\mu$m, where ${L_2} = 1/(2{\mathop{\rm Re}\nolimits} {k_2})$ and ${L_3} = 1/(2{\mathop{\rm Re}\nolimits} {k_3})$  by numerically mode-solving the Maxwell equation. Here, we assumed  $\lambda $=1550 nm,  $t$=5 nm and  ${\varepsilon _3}$=2.5. The experimental data for the permittivity of gold \cite{Johnson1972} is used as  ${\varepsilon_1}$. Mode extension is the width where the intensity decreases to $1/e$ of its maximum value. In the figure a small mode extension on the order of 10 $\mu$m indicates a bound mode and an extremely large mode extension exceeding 1000 $\mu$m a radiation mode. It clearly shows the boundary between the radiation modes and the bound modes in agreement with the analytically predicted cutoff condition Eq.~(\ref{eq6}), (\ref{eq8}) and (\ref{eq9}) (the black curve).

The radiation direction can be switched by a minor deviation of the magnetic field (Fig.~\ref{fig:3}(a)) or the index-asymmetry (Fig.~\ref{fig:2}(b)) from the degenerate cutoff-state $({g_0},{\delta _0})$ which was predicted by Eq.~(\ref{eq9}). Fig.~\ref{fig:4} show the mode distributions for the bound mode (a) and the directional radiation modes in the both sides, which are realized by negative (b) and positive (c) deviations of the gyration from the degenerate cutoff.

Eq.~(\ref{eq6b}) predicts that the minimum gyration ${g_{\deg }}$ needed for the degeneracy is in inverse proportion to the wavelength $\lambda$ (the blue curves of Fig.~\ref{fig:5}(a)) and in proportion to the metal film thickness $t$ (the blue curves of Fig.~\ref{fig:5}(b)). The analytically predicted wavelength- and metal-thickness-dependence of ${g_{\deg }}$  is well agree with the numerical mode solution results (the blue circles of Fig.~\ref{fig:5}). It is noted that Eq.~(\ref{eq6b}) also predicts that ${g_{\deg}}$ can be further degreased by decreasing the environmental permittivity ${\varepsilon_3}$.

We note that the terminology 'degeneracy' here should not be confused with the degeneracy of energy eigenstates in quantum mechanics. The magnetically-induced degeneracy of cutoff-states is lifted by removing the external magnetic field albeit the degeneracy of energy eigenstates in quantum mechanics is lifted by introducing an external magnetic field (Zeeman effect).

In conclusion we derived the analytical cutoff condition of long-range surface magneto-plasmon   along thin metal film surrounded by a magneto-optic material on one side and by a nonmagnetic dielectric on the other side. It predicts the magnetically-induced degenerate cutoff-states between the radiation modes in the magneto-optic layer and those in the nonmagnetic dielectric. Near the degenerate cutoff the radiation modes in one side layer can be switched to those in the other by a minor variation of the magnetic field. The magnetization needed for the degeneracy is proportional to the metal film thickness and inversely proportional to the wavelength. The analytical prediction well agrees with the numerical mode solution results. The predicted new phenomena could find applications in directional nanoantenna and optical switching as well as in sensing technology.
\bibliography{magnet}
\end{document}